\RequirePackage{fix-cm}
\documentclass[smallextended]{svjour3}       
\smartqed  
\usepackage{graphicx}
\usepackage{mathtools}
\usepackage{cryptocode}
\usepackage{underscore}
\usepackage{amssymb}
\begin{document}

\title{Cryptanalysis of Khatoon et al.'s ECC-based Authentication Protocol for Healthcare Systems
}


\author{Mahdi Nikooghadam         \and
        Haleh Amintoosi* 
}


\institute{Faculty of Engineering, Ferdowsi University of Mashhad, Mashhad, Iran\\
              \email{mahdi.nikooghadam@mail.um.ac.ir}           
           \and
           \email{amintoosi@um.ac.ir}
}

\date{Received: date / Accepted: date}

\maketitle

\begin{abstract}

Telecare medical information systems are gaining rapid popularity in terms of providing the delivery of online health-related services such as online remote health profile access for patients and doctors. Due to being installed entirely on Internet, these systems are exposed to various security and privacy threats. Hence, establishing a secure key agreement and authentication process between the patients and the medical servers is an important challenge. Recently, Khatoon et.al proposed an ECC-based unlink-able authentication and key agreement method for healthcare related application in smart city. In this article, we provide a descriptive analysis on their proposed scheme and prove that Khatoon et al.'s scheme is vulnerable to known-session-specific temporary information attack and is not able to provide perfect forward secrecy.

\end{abstract}

\section{Introduction}
\label{intro}
With recent advances in information technology, we are facing a growth in the development of healthcare related applications, such as telecare medical information systems (TMISs) which have been established to provide online healthcare services for patients. In such systems, the patients' medical information such as blood pressure are stored in medical databases. In order to make use of remote health care related services, the patient has to register with the TMIS medical server. To provide service for the patient, the first step is to verify the legitimacy of the patient by the medical server. If patient's legitimacy is verified, healthcare staff and/or doctors are contacted to provide him with the required healthcare consultation.

Despite all the benefits, establishing a secure and privacy-aware communication between the patient and the server is still a major challenge. Failure to provide secure communication may enable the adversary to obtain unauthorized access to the patients' private health data or inject falsified data into the system, resulting in false diagnosis or injury. Hence, research has recently focused on providing secure authentication and communication schemes for TMISs~\cite{Ostad-Sharif2019,Ravanbakhsh2018,Chaudhry2018,Safkhani2019,Jiang2018,Khatoon2019}.

Recently, Khatoon et al.~\cite{Khatoon2019} proposed an anonymous and mutual key agreement scheme based on Elliptic Curve Cryptography (ECC) for TMIS and claimed that their protocol withstands various attacks and satisfies the basic security requirements such as anonymity and un-linkability. In  this paper, we show that their proposed scheme is vulnerable to known-session-specific temporary information attack and does not provide perfect forward secrecy.
\begin{table}
\caption{Notations used in Khatoon et al.'s scheme~\cite{Khatoon2019}}
\label{tab:KhN}       
\begin{tabular}{ll}
\hline\noalign{\smallskip}
symbol & description \\
\hline\noalign{\smallskip}
$q$ & a large prime \\
$e$ & a bilinear map e: $G_1 \times G_1 \rightarrow G_2$ \\
$P$ & The generator of $G_1$ \\
$ID_i, PW_i, B_i$ & Patient's identity, password and biometric information \\
$S$ & TMIS server \\
$s$ & Master private key $s \in Z_q^*$ of $S$ \\
$P_{pub}$ & Public key $P_{pub}=sP$ of $S$ \\
$h$ & A hash function $h: \{0,1\}^* \rightarrow Z_q$ \\
$H$ & A hash function $H_1: \{0,1\}^* \rightarrow G_1$ \\
$T_{k_i}$ & Encryption with symmetric key $k_i$  \\
$T_s , T_i$ & Time stamp of $U_i$ and $S$ \\
\noalign{\smallskip}\hline\noalign{\smallskip}
\end{tabular}
\end{table}

\begin{figure}
\procedure{Registration Phase}{%
\textbf{patient $ U_i$} \> \> \textbf{TMIS server S}  \\
\text{Computes $C_i=PW^i \oplus H_B(B_i)$}\\
 \> \sendmessageright{top=\text{$C_i,ID_i$} , bottom=\text{(Secure Channel)}} \> \\
\> \>
\text{S checks the  $ID_i$ in its database} \\
\> \>\text{if new, S records N=0}\\
\> \>\text{otherwise S records N=N+1}\\
\> \> \text{Computes $V_i=h(ID_i||C_i)$ and $W_i=C_i \oplus h(ID_i||s)$}\\
\> \> \text{Customizes $SC_i$ with($V_i,W_i,P_{pup},h,H,H_B$)} \\
\> \> \text{sends it securely to $U_i$} \\
}
\end{figure}
\begin{figure}
\procedure{Login and Authentication Phase}{
\textbf{Patient $ U_i$} \> \> \textbf{TMIS server S}  \\
\text{$U_i$ insert his smart card $SC_i$ in card reader} \\
\text{Input $ID_i,PW_i$ and imprints $B_i$}\\
\text{the $SC_i$ computes $h(ID_i||PW_i \oplus H_B(B_i))$}\\
\text{Checks $h(ID_i||PW_i \oplus H_B(B_i))=V_i$}\\
\text {if invalid, $SC_i$ aborts the session}\\
\text{Otherwise,}\\
\text{selects $r_i \in Z_p$ and fresh $T_i$}\\
\text{Computes $Q_i=H(ID_i),Q_s=H(ID_s)$}\\
\text{$R_i=r_i.Q_i,K_i=e(P_{pub},r_i.Q_s)$}\\
\text{$Auth_i=E_{k_i}(ID_i||T_i||r_i)$}\\
\> \sendmessageright{top=\text{${R_i,T_i,Auth_i}$}} \> \\
\> \>
\text{Upon receiving $LR_i,S$ checks} \\
\> \> \text{ $\Delta T< T_s-T_i$ if valid it proceed} \\
\> \> \text{And calculate $K_s=e(s,R_i.P)$} \\
\> \> \text{decrypts $Auth_i$ to obtain $(ID_i||T_i||r_i)$ } \\
\> \> \text {Computes $Q_i=H(ID_i)$}\\
\> \> \text{checks $R_i=r_i.Q_i.$ If valid} \\
\> \> \text{Then S generates a random number $r_s$} \\
\> \> \text{Computes $Q_s=H(ID_s), R_s=r_s.Q_i , L_s=r_s.R_i$}\\
\> \> \text{$Auth_s=h(T_i||R_i||T_s||R_s||L_s||K_s)$}\\
\> \> \text{$SK_s=h(T_i||R_i||T_s||R_s||L_s)$}\\
\> \sendmessageleft{top=\text{$R_s,T_s,Auth_s$} , bottom=\text{}} \> \\
\text{verifies $\Delta T<T_s-T_i$} \\
\text{ if valid} \\
\text{Computes $L_i=r_i.R_s$} \\
\text{verifies $Auth_s=h(T_i||R_i||T_s||R_s||L_s||K_i)$}\\
\text{And computes $SK_i=h(T_i||R_i||T_s||R_s||L_s)$}\\
}
\caption{Registration and Authentication phase of Khatoon et al.'s scheme~\cite{Khatoon2019}}
\label{fig:KhRegAut}
\end{figure}

\section{Overview and Cryptanalysis of Khatoon et al.'s Scheme}
In this section, we review and analyse Khatoon et al.'s scheme~\cite{Khatoon2019} and show that it suffers from known-session-specific temporary information attack and cannot guarantee perfect forward secrecy.

\subsection{Overview of Khatoon et al.'s Scheme}
The notations used in Khatoon et al.'s scheme are shown in Table~\ref{tab:KhN}. Their proposed protocol is also demonstrated in Figure~\ref{fig:KhRegAut}.
Before accessing the medical server services, the patient has to register to the server. To do so, the medical server sends the required log in information to the patient through the registration phase. Once registration is done, the patient is able to share a key with the server via the authentication phase. The shared key can then be used for their subsequent secure communications.

\subsection{Cryptanalysis of Khatoon et al.'s Scheme}
In this section, we first demonstrate that the scheme proposed by Khatoon et al.~\cite{Khatoon2019} suffers from the known-session-specific temporary information attack and then, show that it is not able to provide perfect forward secrecy.
\subsubsection{Vulnerability to Known-session-specific Temporary Information Attack}
As mentioned in~\cite{Ostad-Sharif2019}, known-session-specific temporary information attack occurs when the adversary is successful in obtaining the session key by knowing the session random numbers.
In the following, we demonstrate that Khatoon et al.'s scheme is vulnerable to known-session-specific temporary information attack.
\begin{itemize}
\item  As mentioned in the authentication step of Khatoon et al.'s scheme in Figure~\ref{fig:KhRegAut}, $R_i$ is exchanged on public channel, so, the adversary is able to obtain it. Also, $r_s$ is a random parameter which is supposed to be accessible by the adversary in known-session-specific temporary information attack. Hence, the adversary is able to compute $L_s$ as $L_s=r_s.R_i$.
\item As shown in Figure~\ref{fig:KhRegAut}, the session key $SK$ is computed as $SK_i=h(T_i||R_i||T_s||R_s||L_s)$. Parameters $R_i, T_i, R_s, T_s$ are exchanged on public channel, so they are available to the adversary. As stated above, the adversary is able to compute $L_s$. Having all the parameters included in $SK$, he is now able to compute the session key $SK$. This clearly states that Khatoon et al.'s scheme is prone to known-session-specific temporary information attack.
\end{itemize}

\subsubsection{Perfect Forward Insecurity}
The protocol is said to provide perfect forward secrecy if, by knowing the longterms such as the server's public/private keys, the adversary  is not able to compute the session key $SK$. In the following, we show that Khatoon et al.'s scheme does not guarantee perfect forward secrecy.

\begin{itemize}
\item  Lets assume that the adversary knows the medical server's public and private keys. So, he is able to compute $K_s$ as $K_s=e(s,R_i.P)$, since $R_i$ is available on public channel.

\item As $K_s = K_i$, he is now able to decrypt $Auth_i$ and obtain $r_i$ as $Auth_i=E_{k_i}(ID_i||T_i||r_i)$.
\item
    Having $r_i$ and $R_i$ on public channel, the adversary computes $L_i=r_i.R_s$.

\item  On the other hand, $L_i = r_i.R_s=r_i.r_s.Q_i=r_s.r_i.Q_i=r_s.R_i=L_s$. So, the adversary already has $L_s$ at hand too.

\item Having $L_s$ computed in the above step and having access to $T_i, R_i, T_s$ and $R_s$ on public channel, the adversary is now able to compute the session key $SK$ as $SK_i=h(T_i||R_i||T_s||R_s||L_s)$. This means that Khatoon et al.'s scheme is not able to provide perfect forward secrecy.
\end{itemize}

\section{Conclusion and Future Work}
Providing a secure and privacy-preserving communication channel between different patients and medical systems in remote healthcare systems has gain lots of attention. In this article, we reviewed the authentication and key agreement protocol presented by Khatoon et al., and demonstrated that it is prone to known-session-specific temporary information attack and does not provide perfect forward secrecy. In future, we plan to present a secure and privacy preserving registration and key agreement scheme for healthcare systems that addresses the shortcomings of related work.



\begin{thebibliography}{}
\bibitem{Ostad-Sharif2019}
Ostad‐Sharif, A, Abbasinezhad‐Mood, D, Nikooghadam, M. An enhanced anonymous and unlinkable user authentication and key agreement protocol for TMIS by utilization of ECC. Int J Commun Syst. 32:e3913. https://doi.org/10.1002/dac.3913, (2019)
\bibitem{Ravanbakhsh2018}
Ravanbakhsh N, Nazari M. An efficient improvement remote user mutual authentication and session key agreement scheme for E-healthcare systems. Multimed Tools Appl. vol. 77, no. 1, pp. 55‐88, (2018)

\bibitem{Chaudhry2018}
Chaudhry, S.A., Naqvi, H. , Khan, M.K., An enhanced lightweight anonymous biometric based authentication scheme for TMIS, Multimed Tools Appl vol. 77, no. 5, : 5503-5524. (2019)

\bibitem{Safkhani2019}
M. Safkhani and A. Vasilakos, A New Secure Authentication Protocol for Telecare Medicine Information System and Smart Campus, IEEE Access, vol. 7, pp. 23514-23526, (2019)

\bibitem{Jiang2018}
Jiang, Q., Chen, Z., Li, B. et al. Security analysis and improvement of bio-hashing based three-factor authentication scheme for telecare medical information systems, J Ambient Intell Human Comput, vol. 9, no. 4, pp: 1061-1073, (2018)
\bibitem{Khatoon2019}
S. Khatoon, S. M. M. Rahman, M. Alrubaian and A. Alamri, "Privacy-Preserved, Provable Secure, Mutually Authenticated Key Agreement Protocol for Healthcare in a Smart City Environment," in IEEE Access, vol. 7, pp. 47962-47971, (2019)

\end{thebibliography}
\end{document}